\documentclass[english,aps,superscriptaddress]{revtex4}
\usepackage[T1]{fontenc}
\usepackage[latin9]{inputenc}
\usepackage{mathrsfs}
\usepackage{amsmath}
\usepackage{amssymb}
\usepackage{amsbsy}
\usepackage{esint}
\makeatletter
\@ifundefined{textcolor}{}
{%
 \definecolor{BLACK}{gray}{0}
 \definecolor{WHITE}{gray}{1}
 \definecolor{RED}{rgb}{1,0,0}
 \definecolor{GREEN}{rgb}{0,1,0}
 \definecolor{BLUE}{rgb}{0,0,1}
 \definecolor{CYAN}{cmyk}{1,0,0,0}
 \definecolor{MAGENTA}{cmyk}{0,1,0,0}
 \definecolor{YELLOW}{cmyk}{0,0,1,0}
}

\makeatother

\usepackage{babel}

\begin{document}


\title{Disentangling covariant Wigner functions for chiral fermions}

\author{Jian-Hua Gao}
\affiliation{Shandong Provincial Key Laboratory of Optical Astronomy and Solar-Terrestrial
Environment, Institute of Space Sciences, Shandong University, Weihai,
Shandong 264209, China}

\author{Zuo-Tang Liang}
\affiliation{School of Physics and Key Laboratory of Particle Physics and Particle
Irradiation (MOE), Shandong University, Jinan, Shandong 250100, China}

\author{Qun Wang}
\affiliation{Department of Modern Physics, University of Science and Technology
of China, Hefei, Anhui 230026, China}

\author{Xin-Nian Wang}
\affiliation{Key Laboratory of Quark and Lepton Physics (MOE) and Institute of
Particle Physics, Central China Normal University, Wuhan, 430079, China}

\affiliation{Nuclear Science Division, MS 70R0319, Lawrence Berkeley National
Laboratory, Berkeley, California 94720}

\begin{abstract}
We develop a general formalism for the quantum kinetics of chiral fermions in a background electromagnetic field
based on a semiclassical expansion of covariant Wigner functions in the Planck constant $\hbar$.
We demonstrate to any order of $\hbar$ that only the time-component of the Wigner function is independent
while other components are explicit derivative. We further demonstrate to any order of $\hbar$
that a system of quantum kinetic equations for multiple-components of Wigner functions
can be reduced to one chiral kinetic equation involving only the single-component distribution function.
These are remarkable properties of the quantum kinetics of chiral fermions and
will significantly simplify the description and simulation of chiral effects
in heavy ion collisions and Dirac/Weyl semimetals.
We present the unintegrated chiral kinetic equations in four-momenta up to $O(\hbar ^2)$ and
the integrated ones in three-momenta up to $O(\hbar)$.
We find that some singular terms emerge in the integration over the time component of
the four-momentum, which result in a new source term contributing to the chiral anomaly,
in contrast to the well-known scenario of the Berry phase term.
Finally we  rewrite our results in any Lorentz frame with a reference four-velocity
and show  how  the non-trivial transformation of the distribution function in different frames emerges 
in a natural way.

\end{abstract}

\maketitle

\section{Introduction}

Recently the properties of chiral fermions in electromagnetic fields
have been extensively studied in high energy heavy ion collisions
\cite{Kharzeev:2012ph,Kharzeev:2015znc} as well as in Dirac or Weyl
semi-metals \cite{Son:2012bg,Li:2014bha,PhysRevX.5.031023}. One of
the most important effects for chiral fermions is the Chiral Magnetic
Effect (CME) \cite{Vilenkin:1980fu,Kharzeev:2007jp,Fukushima:2008xe,Son:2009tf,Kharzeev:2010gr,Hou:2011ze}
(for reviews, see, e.g., Ref. \cite{Kharzeev:2012ph,Kharzeev:2015znc}).
The CME is closely related to the chiral anomaly and topological structure
of gauge fields \cite{Adler:1969gk,Bell:1969ts,Nielsen:1983rb}. Any
imbalance in the number of right-handed and left-handed quarks and
antiquarks due to topological charge fluctuations of gauge fields
may induce an electric current along the magnetic field which leads
to a Charge Separation Effect (CSE). Though it is very challenging to pin
down the CME or its consequences such as the CSE in heavy ion collisions
\cite{Abelev:2009ac,Abelev:2009ad,Abelev:2012pa,Sorensen:2017},
the CME has recently been confirmed in Dirac or Weyl semi-metals
\cite{Son:2012bg,Li:2014bha,PhysRevX.5.031023}.
The chiral vortical effect (CVE)
\cite{Vilenkin:1978hb,Erdmenger:2008rm,Banerjee:2008th,Son:2009tf,Landsteiner:2011cp,Gao:2012ix}
is another phenomenon for chiral fermions in a fluid,
where the vorticity can be regarded as the local ortibal angular momentum
and can lead to the polarization of particles through spin-orbit
couplings \cite{Liang:2004ph,Liang:2004xn}. The polarization of $\varLambda$
hyperons has been measured recently for the first time in the STAR experiment at Relativistic
Heavy Ion Collider (RHIC) \cite{STAR:2017ckg}.

The kinetic theory is an important tool to describe these novel properties
of chiral fermions in phase space \cite{Son:2012wh,Stephanov:2012ki,Son:2012zy,Chen:2013iga,Gorbar:2016ygi}.
The covariant Wigner function is a powerful and systematic quantum
kinetic approach \cite{Heinz:1983nx,Elze:1986qd,Vasak:1987um,Zhuang:1995pd,Florkowski:1995ei,Blaizot:2001nr,Wang:2001dm}:
it has been shown that the CME, CVE and Covariant
Chiral Kinetic Equation (CCKE) can be derived from covariant Wigner
functions \cite{Gao:2012ix,Chen:2012ca,Gao:2015zka,Gao:2017gfq,Hidaka:2016yjf}.
The Wigner functions have multiple components which are entangled with each other,
while in the Boltzmann-like equation the single-component distribution function is involved.
It is unknown to what extent that a fermionic quantum system can be described by
the Boltzmann-like distribution function.

In this paper we will develop a semiclassical expansion of the covariant Wigner function
in the Planck constant $\hbar$. This expansion is very general and
does not require quasi-equilibrium conditions, so it is very different from
the expansion in space-time gradients and field strengths near equilibrium
\cite{Gao:2012ix,Chen:2012ca,Gao:2015zka,Gao:2017gfq}.
In this formalism, we can derive the quantum kinetic equations
for the covariant Wigner function order by order.
We will show how to disentangle the covariant Wigner function
and reduce a system of quantum kinetic equations
for multiple components of the covariant Wigner function
to one chiral kinetic equation (CKE) involving
only the single-component distribution function.
We will also present other properties of the CKE and the chiral anomaly.
Finally we will generalize our results to any Lorentz frame
with a reference four-velocity. We will show that
the side-jump effect naturally emerges from the change of
the first order distribution function
when one chooses a different four-velocity.

We use the same sign convention for the fermion charge $Q$ as in
Refs. \cite{Vasak:1987um,Gao:2012ix,Chen:2012ca,Gao:2015zka,Gao:2017gfq}.
The sign convention for the axial vector component of the Wigner function
is the same as in Refs. \cite{Gao:2012ix,Chen:2012ca,Gao:2015zka,Gao:2017gfq}
but opposite to Ref. \cite{Vasak:1987um}.

\section{Covariant Wigner functions}

In a background electromagnetic field, the quantum anologue of a classical
phase-space distribution for fermions is the covariant Wigner function
\cite{Heinz:1983nx,Elze:1986qd,Vasak:1987um},
\begin{equation}
W(x,p)=\left\langle :\hat{W}(x,p):\right\rangle ,
\end{equation}
where $\hat{W}_{\alpha\beta}(x,p)$ is the Wigner operator, the brackets
denote the ensemble average and the colons the normal ordering of
the operators, and $x$ and $p$ are the space-time and four-momentum vector respectively.
The Wigner operator is defined by
\begin{equation}
\hat{W}_{\alpha\beta}(x,p)=\int\frac{d^{4}y}{(2\pi)^{4}}e^{-ip\cdot y}\bar{\psi}_{\beta}\left(x+\frac{1}{2}y\right)U\left(x+\frac{1}{2}y,x-\frac{1}{2}y\right)\psi_{\alpha}\left(x-\frac{1}{2}y\right),\label{eq:winger_02}
\end{equation}
where $\psi_{\alpha}$ and $\bar\psi_{\beta}$ are Dirac spinors with
$\alpha,\beta$ being the spinor indices running from 1 to 4, the
gauge link $U$ is defined as $U(x_{1},x_{2})=\exp\left[-iQ\int_{x_{2}}^{x_{1}}dz_\mu A^\mu (z)\right]$
along a straight path between the points $x_1$ and $x_2$
with $A^\mu (z)$ being the vector potential of the background electromagnetic field.
The Wigner function can be decomposed in 16 independent generators of Clifford algebra,
\begin{equation}
W=\frac{1}{4}\left[\mathscr{F}+i\gamma^{5}\mathscr{P}+\gamma^{\mu}\mathscr{V}_{\mu}+\gamma^{5}\gamma^{\mu}\mathscr{A}_{\mu}+\frac{1}{2}\sigma^{\mu\nu}\mathscr{S}_{\mu\nu}\right],\label{eq:wigner-decomp}
\end{equation}
whose coefficients $\mathscr{F}$, $\mathscr{P}$, $\mathscr{V}_{\mu}$,
$\mathscr{A}_{\mu}$ and $\mathscr{S}_{\mu\nu}$ are the scalar, pseudo-scalar,
vector, axial-vector and tensor components of the Wigner function respectively.
For massless fermions, $\mathscr{V}_{\mu}$ and $\mathscr{A}_{\mu}$
are decoupled from the rest components which can be linearly combined
into the vector component of the covariant Wigner function
with chirality (VWC) \cite{Gao:2012ix,Gao:2015zka},
\begin{equation}
\mathscr{J}_{\mu}^{s}(x,p)=\frac{1}{2}[\mathscr{V}_{\mu}(x,p)+s\mathscr{A}_{\mu}(x,p)]\,,
\label{vwfc}
\end{equation}
where $s=\pm$ is the chirality and $\mu =0,1,2,3$ denotes the Lorentz indices.
These components satisfy following equations (see Eqs. (5.12-5.21) of Ref. \cite{Vasak:1987um})
\begin{eqnarray}
\Pi^{\mu}\mathscr{J}_{\mu}^{s}(x,p) & = & 0,\nonumber \\
G^{\mu}\mathscr{J}_{\mu}^{s}(x,p) & = & 0,\nonumber \\
2s(\Pi^{\mu}\mathscr{J}_{s}^{\nu}-\Pi^{\nu}\mathscr{J}_{s}^{\mu}) & = & -\hbar\epsilon^{\mu\nu\rho\sigma}G_{\rho}\mathscr{J}_{\sigma}^{s}\,,
\label{eq:wig-eq-1}
\end{eqnarray}
where $\epsilon^{\mu\nu\sigma\beta}$ is the anti-symmetric tensor and the  operators $\Pi^{\mu}$ and $G^{\mu}$ are defined by
\begin{eqnarray}
\Pi^{\mu} & = & p^{\mu}-\hbar\frac{1}{2}j_{1}(z) QF^{\mu\nu}\partial_{\nu}^{p},\nonumber \\
G^{\mu} & = & \partial_{x}^{\mu}-j_{0}(z) QF^{\mu\nu}\partial_{\nu}^{p},\label{eq:op}
\end{eqnarray}
where $z\equiv \hbar\Delta /2$ with $\Delta \equiv \partial_{x}\cdot\partial_{p}$ being the differential operator,
and $j_{0}(z)=\sin z/z$ and $j_{1}(z)=(\sin z-z\cos z)/z^{2}$ are spherical Bessel functions.
Note that $\partial_{x}$ in the operator $\Delta$ acts only on $F^{\mu\nu}$ to its right but not on other functions.
We see that we have recovered the $\hbar$ dependence explicitly in Eqs.(\ref{eq:wig-eq-1}) and (\ref{eq:op})
in order to perform the $\hbar$ expansion.

We see that each equation in (\ref{eq:wig-eq-1}) is a polynomial
in the Planck constant $\hbar$. So we can make a semiclassical expansion
in powers of $\hbar$ for $\mathscr{J}_{\mu}$ and the operators $\Pi^{\mu}$ and $G^{\mu}$.
We note that the idea of the semiclassical expansion in $\hbar$ for
the Wigner functions in electromagnetic fields was first proposed
in Ref. \cite{Vasak:1987um} but without concrete calculations.
It was recently used in Refs. \cite{Son:2012zy,Manuel:2013zaa,Manuel:2014dza,Hidaka:2016yjf,Huang:2018wdl}
for chiral fermions at $O(\hbar)$.
The expansions of $\Pi^{\mu}$ and $G^{\mu}$ are
\begin{equation}
\mathscr{J}_{\mu} = \sum _{n=0}^\infty \hbar^{n}\mathscr{J}^{(n)}_{\mu},\,\,\,
\Pi ^{\mu}=\sum _{n=0}^\infty \hbar^{2n} \Pi ^\mu _{(2n)},\,\,\,
G^{\mu}=\sum _{n=0}^\infty \hbar^{2n} G^\mu _{(2n)}\,,
\label{eq:expansion}
\end{equation}
where $n$ are nonnegative integers and we have suppressed the helicity
indices of $\mathscr{J}_{\mu}^{s}(x,p)$ for simplicity of notations.
We see that the operator $\Pi ^{\mu}$ and $G^{\mu}$ have only even-order terms
and can be put into compact forms ($n\neq 0$)
\begin{eqnarray}
\Pi_{(2n)}^{\mu} & = & \frac{(-1)^{n}n}{2^{2n-1}(2n+1)!}
\Delta ^{2n-1} F^{\mu\nu}\partial_{\nu}^{p}, \nonumber\\
G_{(2n)}^{\mu} & = & \frac{(-1)^{n+1}}{2^{2n}(2n+1)!}\Delta ^{2n}
F^{\mu\nu}\partial_{\nu}^{p} \,.
\label{g-pi-n}
\end{eqnarray}
Up to the second order these operators have the form
\begin{eqnarray}
\Pi^{\mu} & = & \Pi^{\mu}_{(0)}+\Pi^{\mu}_{(2)} \equiv  p^{\mu}-\frac{Q}{12}\hbar^2 \Delta F^{\mu\nu}\partial_\nu^p \,, \nonumber \\
G^{\mu} & = & G^{\mu}_{(0)} + G^{\mu}_{(2)} \equiv  \partial_{x}^{\mu}-QF^{\mu\nu}\partial_{\nu}^{p}+\frac{Q}{24}\hbar^2 \Delta^2 F^{\mu\nu}\partial_\nu^p \,.
\label{op-2nd}
\end{eqnarray}
We can write these operators explicitly in time and space components:
$\Pi^{\mu}=(\Pi_0,\mathbf{\Pi})$, $G^{\mu}=(G_0,-\mathbf{G})$,
whose expressions in each order are displayed in Eq. (\ref{op-explicit}).

\section{Semiclassical expansion: a general formalism}
\label{semi-class}

In this section we make a semiclassical expansion
in powers of $\hbar$ for Eq. (\ref{eq:wig-eq-1})
using the expansions of $\mathscr{J}_{\mu}$, $\Pi^{\mu}$ and $G^{\mu}$
in Eqs. (\ref{eq:expansion},\ref{g-pi-n}).

Let us write Eq. (\ref{eq:wig-eq-1}) explicitly in time and spatial components
\begin{eqnarray}
{\Pi}_{0}\mathscr{J}_{0}-\mathbf{\Pi}\cdot\pmb{\mathscr{J}} & = & 0,
\label{eq:constraint-1}\\
{G}_0\mathscr{J}_{0}+\mathbf{G}\cdot\pmb{\mathscr{J}} & = & 0,
\label{eq:evo-1}\\
\hbar\left[{G}_0  \pmb{\mathscr{J}} + \mathbf{G} \mathscr{J}_{0}\right] & = & 2s(\mathbf{\Pi}\times\pmb{\mathscr{J}})\,,
\label{eq:evo-2} \\
-\hbar \mathbf{G} \times \pmb {\mathscr{J}}  & = & 2s\left(\mathbf{\Pi} \mathscr{J}_{0}-\Pi_0   \pmb{\mathscr{J}} \right) \,,
\label{eq:constraint-2}
\end{eqnarray}
where we have used $\mathscr{J}^{\mu}=(\mathscr{J}_{0},\pmb{\mathscr{J}})$.
We note that Eqs. (\ref{eq:evo-2},\ref{eq:constraint-2}) come from the third line of Eq. (\ref{eq:wig-eq-1}).
These equations can be grouped into evolution equations
with the operator $G_0$ which has the time derivative, Eqs. (\ref{eq:evo-1},\ref{eq:evo-2}),
and constraint equations without it, Eqs. (\ref{eq:constraint-1},\ref{eq:constraint-2}).

Inserting Eq. (\ref{eq:expansion}) into Eqs. (\ref{eq:constraint-1}-\ref{eq:constraint-2})
and using Eqs. (\ref{g-pi-n},\ref{op-2nd}), we obtain a system of quantum kinetic equations at any power of $\hbar$.
The evolution equation (\ref{eq:evo-1}) at $O(\hbar ^n)$ and (\ref{eq:evo-2}) at $O(\hbar ^{n+1})$ read
\begin{eqnarray}
\sum_{i=0}^{[n/2]}\left[G_{0}^{(2i)}\mathscr{J}_{0}^{(n-2i)}
+\mathbf{G}^{(2i)}\cdot\pmb{\mathscr{J}}^{(n-2i)}\right] & = & 0\,,
\label{evo-j0}\\
\sum_{i=0}^{[n/2]} \left[ G_0^{(2i)}\pmb{\mathscr{J}}^{(n-2i)} + {\mathbf G}^{(2i)} {\mathscr{J}}_0^{(n-2i)}\right]
&=& 2s\sum_{i=0}^{[(n+1)/2]} {\mathbf \Pi}^{(2i)}\times \pmb {\mathscr{J}}^{(n-2i+1)}\,,
\label{evo-jv}
\end{eqnarray}
where $[n/2]$ denotes the largest integer bounded by $n/2$.
We emphasize again that Eq. (\ref{evo-j0}) and (\ref{evo-jv}) are called the evolution equations
due to the fact that the former contains the time derivative term $\partial_t \mathscr{J}_{0}^{(n)}$
while the latter contains $\partial_t\pmb{\mathscr{J}}^{(n)}$.
In Eq. (\ref{evo-jv}) we see that the time derivative term $G_0^{(0)}\pmb{\mathscr{J}}^{(n)}$
is on the left-hand side while the $(n+1)$-th order term
$\mathbf{p} \times \pmb {\mathscr{J}}^{(n+1)}$ is on the right-hand side,
so it seems to be impossible to obtain the $n$-th order solution $\pmb{\mathscr{J}}^{(n)}$
without knowing the $(n+1)$-th order one, but we will show that it is possible
due to some good properties of $\mathscr{J}^\mu$.
The constraint equation (\ref{eq:constraint-1}) and (\ref{eq:constraint-2})
at $O(\hbar ^n)$ and $O(\hbar ^{n+1})$ respectively read
\begin{eqnarray}
\sum_{i=0}^{[n/2]}\left[\Pi_{0}^{(2i)}\mathscr{J}_{0}^{(n-2i)}-\boldsymbol{\Pi}^{(2i)}\cdot\pmb{\mathscr{J}}^{(n-2i)}\right] & = & 0\,,
\label{constr-J0} \\
2s \sum_{i=0}^{[(n+1)/2]} \left[{\mathbf \Pi}^{(2i)} {\mathscr{J}}_0^{(n-2i+1)}
-\Pi _0^{(2i)} \pmb{\mathscr{J}}^{(n-2i+1)} \right] &=&
-\sum_{i=0}^{[n/2]} {\mathbf G}^{(2i)}\times \pmb{\mathscr{J}}^{(n-2i)}\,.
\label{constr-Jv}
\end{eqnarray}
We note that Eq. (\ref{evo-jv}) and (\ref{constr-Jv})
come from Eq. (\ref{eq:evo-2}) and (\ref{eq:constraint-2}) respectively
which relate the $(n+1)$-th order terms on the right-hand side to
the $n$-th order terms on the left-hand side.
Equation (\ref{constr-J0}) leads to mass-shell conditions from which
one can determine the dispersion relations of chiral fermions.

From Eq. (\ref{constr-Jv}) we can solve $\pmb{\mathscr{J}}^{(n+1)}$ as
\begin{eqnarray}
\pmb{\mathscr{J}}^{(n+1)} & = & \frac{\mathbf{p}}{p_{0}}\mathscr{J}_{0}^{(n+1)}
+\frac{s}{2p_{0}} \sum_{i=0}^{[n/2]}\mathbf{G}^{(2i)}\times\pmb{\mathscr{J}}^{(n-2i)} \nonumber\\
 &  & +\frac{1}{p_{0}}\sum_{i=1}^{[(n+1)/2]}\left[\boldsymbol{\Pi}^{(2i)}\mathscr{J}_{0}^{(n-2i+1)}-\Pi ^{(2i)}_{0}\pmb{\mathscr{J}}^{(n-2i+1)}\right] \,.
\label{general-solution}
\end{eqnarray}
We see that on the right-hand side all $\pmb{\mathscr{J}}$ terms are in lower order than $\pmb{\mathscr{J}}^{(n+1)}$,
and the only $(n+1)$-th order term is $(\mathbf{p}/p_{0})\mathscr{J}_{0}^{(n+1)}$.
Equation (\ref{general-solution}) also holds for $n=-1$ which gives
$\pmb{\mathscr{J}}^{(0)}=(\mathbf{p}/p_{0})\mathscr{J}_{0}^{(0)}$.


By recursively applying Eq. (\ref{general-solution}) to all lower order $\pmb{\mathscr{J}}$, we can finally
express $\pmb{\mathscr{J}}^{(n+1)}$ in terms of $\mathscr{J}_0^{(i)}$ for $i=1,2,\cdots ,n+1$. When substituting
Eq. (\ref{general-solution}) into Eq. (\ref{evo-jv}), the highest order term $\mathbf{p} \times \pmb{\mathscr{J}}^{(n+1)}$
on the right-hand side of Eq. (\ref{evo-jv}) is vanishing due to the term $(\mathbf{p}/p_{0})\mathscr{J}_{0}^{(n+1)}$
in $\pmb{\mathscr{J}}^{(n+1)}$. Therefore the evolution equation (\ref{evo-jv}) for $\pmb{\mathscr{J}}^{(n)}$
can be finally converted to an equation for $\mathscr{J}_0^{(i)}$ ($i=1,2,\cdots ,n$)
with a term of $\sim\partial _t \mathscr{J}_{0}^{(n)}$ at the highest order,
therefore Eq. (\ref{evo-jv}) is closed and can be considered as another evolution equation
for $\mathscr{J}_{0}^{(n)}$ besides Eq. (\ref{evo-j0}).
We call this equation the derived evolution equation for $\mathscr{J}_{0}^{(n)}$ as it is from
that for $\pmb{\mathscr{J}}^{(n)}$, while we call Eq. (\ref{evo-j0}) the original one.
Then the question arises: is the derived evolution equation consistent with the original one?
The answer is positive for any order $n$. By using Eq. (\ref{general-solution}) and mathematical induction,
we can prove that Eq. (\ref{evo-jv}) is automatically satisfied for any $n$
once the evolution equation (\ref{evo-j0}) and the mass-shell conditions in Eq. (\ref{constr-J0}) are satisfied with Eq. (\ref{general-solution}). This means that to any order of $\hbar$ the evolution equations
for the vector component $\pmb{\mathscr{J}}$ are redundant, the only one that is needed
is the original evolution equation for the time-component $\mathscr{J}_0$ constrained by the mass-shell conditions.
The proof of this statement is given in Appendix \ref{proof-app}.

In summary, we have proved to any order of $\hbar$ that only the time-component of the VWC is independent
while the spatial components can be derived from it explicitly.
We further demonstrate to any order of $\hbar$ that a system of quantum kinetic equations
for multiple-components of Wigner functions can be reduced to
one chiral kinetic equation involving only the single-component distribution function.
These are remarkable properties of the quantum kinetics of chiral fermions,
which, if combined with collision terms \cite{Chen:2015gta,Hidaka:2016yjf}, will
significantly simplify the description and simulation of quantum kinetic evolution for chiral effects
in heavy ion collisions and chiral materials such as Dirac or Weyl semimetals.

\section{Second order results}
\label{2nd}

As an application of the general formalism, in this section we will derive
the chiral kinetic equations and $\mathscr{J}_{0}$ to the second order in $\hbar$.

We write the explicit forms of evolution equations in (\ref{evo-j0}) for $n=0,1,2$ as
\begin{eqnarray}
G_0^{(0)}\mathscr{J}_{0}^{(0)} + \mathbf{G}^{(0)}\cdot\pmb{\mathscr{J}}^{(0)} & = & 0\,,\nonumber \\
G_0^{(0)}\mathscr{J}_{0}^{(1)} + \mathbf{G}^{(0)}\cdot\pmb{\mathscr{J}}^{(1)} & = & 0\,,\nonumber \\
G_0^{(0)}\mathscr{J}_{0}^{(2)} + \mathbf{G}^{(0)}\cdot \pmb{\mathscr{J}}^{(2)} & = &
-G_0^{(2)}\mathscr{J}_{0}^{(0)} - \mathbf{G}^{(2)}\cdot \pmb{\mathscr{J}}^{(0)} \,.
\label{eq:evolution1}
\end{eqnarray}
The explicit forms of evolution equations in (\ref{evo-jv}) for $n=-1,0,1,2$ read
\begin{eqnarray}
0 & = & 2s(\mathbf{p}\times\pmb{\mathscr{J}}^{(0)}),\nonumber \\
G_0^{(0)}\pmb{\mathscr{J}}^{(0)} + \mathbf{G}^{(0)} \mathscr{J}_{0}^{(0)} & = & 2s(\mathbf{p}\times\pmb{\mathscr{J}}^{(1)}),\nonumber\\
G_0^{(0)}\pmb{\mathscr{J}}^{(1)} + \mathbf{G}^{(0)} \mathscr{J}_{0}^{(1)} & = & 2s(\mathbf{p}\times\pmb{\mathscr{J}}^{(2)}
+ \mathbf{\Pi}^{(2)}\times\pmb{\mathscr{J}}^{(0)}),\nonumber\\
G_0^{(0)}\pmb{\mathscr{J}}^{(2)} + \mathbf{G}^{(0)} \mathscr{J}_{0}^{(2)}
+G_0^{(2)}\pmb{\mathscr{J}}^{(0)} + \mathbf{G}^{(2)} \mathscr{J}_{0}^{(0)}
& = & 2s(\mathbf{p}\times\pmb{\mathscr{J}}^{(3)}
+ \mathbf{\Pi}^{(2)}\times\pmb{\mathscr{J}}^{(1)})\,.
\label{eq:evolution2}
\end{eqnarray}
The first line is obtained under the implied assumption
that $\mathscr{J}_0^{(-1)}=0$ and $\pmb{\mathscr{J}}^{(-1)}=\mathbf{0}$.
We see in Eq. (\ref{eq:evolution2}) that the first line is degenerated to a constraint condition for $\pmb{\mathscr{J}}^{(0)}$
and $\mathbf{p}\times\pmb{\mathscr{J}}^{(n+1)}$ appears in the $n$-th order equation.
As we have argued in the previous section that the $n$-th order equation finally involves the same or
lower order quantities. We will see exlicitly that this is really the case.

The constraint equations in (\ref{constr-J0}) for $n=0,1,2$ read
\begin{eqnarray}
p_{0}\mathscr{J}_{0}^{(0)}-\mathbf{p}\cdot\pmb{\mathscr{J}}^{(0)} & = & 0, \nonumber \\
p_{0}\mathscr{J}_{0}^{(1)}-\mathbf{p}\cdot\pmb{\mathscr{J}}^{(1)} & = & 0, \nonumber \\
p_{0}\mathscr{J}_{0}^{(2)}-\mathbf{p}\cdot\pmb{\mathscr{J}}^{(2)}
+ \Pi_{0}^{(2)}\mathscr{J}_{0}^{(0)}-\mathbf{\Pi}^{(2)}\cdot\pmb{\mathscr{J}}^{(0)}& = & 0 \,,
\label{eq:constraints1}
\end{eqnarray}
which provide mass-shell conditions in each order. The constraint equations in (\ref{constr-Jv}) for $n=-1,0,1,2$ read
\begin{eqnarray}
0 & = & 2s(p_{0}\pmb{\mathscr{J}}^{(0)}-\mathbf{p}\mathscr{J}_{0}^{(0)}),\nonumber \\
\mathbf{G}^{(0)}\times\pmb{\mathscr{J}}^{(0)} & = & 2s(p_{0}\pmb{\mathscr{J}}^{(1)}-\mathbf{p}\mathscr{J}_{0}^{(1)}),\nonumber\\
\mathbf{G}^{(0)}\times\pmb{\mathscr{J}}^{(1)} & = &
2s(p_{0}\pmb{\mathscr{J}}^{(2)}-\mathbf{p}\mathscr{J}_{0}^{(2)}+\Pi_{0}^{(2)}\pmb{\mathscr{J}}^{(0)}-\mathbf{\Pi}^{(2)}\mathscr{J}_{0}^{(0)}),\nonumber\\
\mathbf{G}^{(0)}\times\pmb{\mathscr{J}}^{(2)} + \mathbf{G}^{(2)}\times\pmb{\mathscr{J}}^{(0)} & = &
2s(p_{0}\pmb{\mathscr{J}}^{(3)}-\mathbf{p}\mathscr{J}_{0}^{(3)}+\Pi_{0}^{(2)}\pmb{\mathscr{J}}^{(1)}-\mathbf{\Pi}^{(2)}\mathscr{J}_{0}^{(1)})\,.
\label{eq:constraints2}
\end{eqnarray}
From Eq. (\ref{eq:constraints2}), we can express $\pmb{\mathscr{J}}$ in terms of $\mathscr{J}_{0}$ order by order,
\begin{eqnarray}
\pmb{\mathscr{J}}^{(0)} & = & \frac{\mathbf{p}}{p_{0}}\mathscr{J}_{0}^{(0)},\nonumber \\
\pmb{\mathscr{J}}^{(1)} & = & \frac{\mathbf{p}}{p_{0}}\mathscr{J}_{0}^{(1)} + \frac{s}{2p_{0}}\mathbf{G}^{(0)}\times \pmb{\mathscr{J}}^{(0)} ,\nonumber\\
\pmb{\mathscr{J}}^{(2)} & = & \frac{\mathbf{p}}{p_{0}}\mathscr{J}_{0}^{(2)} + \frac{s}{2p_{0}}\mathbf{G}^{(0)}\times \pmb{\mathscr{J}}^{(1)}
- \frac{1}{p_{0}}\Pi_{0}^{(2)}\pmb{\mathscr{J}}^{(0)}+\frac{1}{p_{0}}\mathbf{\Pi}^{(2)}\mathscr{J}_{0}^{(0)},\nonumber\\
\pmb{\mathscr{J}}^{(3)} & = & \frac{\mathbf{p}}{p_{0}}\mathscr{J}_{0}^{(3)}
+ \frac{s}{2p_{0}}\mathbf{G}^{(0)}\times \pmb{\mathscr{J}}^{(2)}
-\frac{1}{p_{0}}\Pi_{0}^{(2)}\pmb{\mathscr{J}}^{(1)}+\frac{1}{p_{0}}\mathbf{\Pi}^{(2)}\mathscr{J}_{0}^{(1)}
+ \frac{s}{2p_{0}}\mathbf{G}^{(2)}\times \pmb{\mathscr{J}}^{(0)}\,.
\label{eq:j0-j}
\end{eqnarray}
We see that $\pmb{\mathscr{J}}^{(1)}$ depends on $\pmb{\mathscr{J}}^{(0)}=(\mathbf{p}/p_{0})\mathscr{J}_{0}^{(0)}$,
so $\pmb{\mathscr{J}}^{(1)}$ is finally determined by $\mathscr{J}_{0}^{(0)}$ and $\mathscr{J}_{0}^{(1)}$.
Similarly $\pmb{\mathscr{J}}^{(2)}$ is determined by $(\mathscr{J}_{0}^{(0)}, \mathscr{J}_{0}^{(1)}, \mathscr{J}_{0}^{(2)})$,
and $\pmb{\mathscr{J}}^{(3)}$ is determined by $(\mathscr{J}_{0}^{(0)}, \mathscr{J}_{0}^{(1)}, \mathscr{J}_{0}^{(2)},\mathscr{J}_{0}^{(3)})$.
So we have explicitly shown that $\pmb{\mathscr{J}}^{(n)}$ is determined by $\mathscr{J}_{0}$ up to the order $n$.
Another property of $\pmb{\mathscr{J}}^{(n)}$ with $n=0,1,2,3$ in Eq. (\ref{eq:j0-j}) is that
the first contribution is proportional to $(\mathbf{p}/p_0)\mathscr{J}_{0}^{(n)}$,
when simplifying $\mathbf{p}\times\pmb{\mathscr{J}}^{(n)}$ in Eq. (\ref{eq:evolution2}) by using Eq. (\ref{eq:j0-j})
the highest order term $(\mathbf{p}/p_0)\mathscr{J}_{0}^{(n)}$ is vanishing.
So each evolution equation of (\ref{eq:evolution2}) for $\mathscr{J}_{0}^{(n)}$ involves only terms of $\mathscr{J}_{0}$ up to the order $n$.
This is a good property of the VWC, with which we can verify that the evolution equations (\ref{eq:evolution2}) for $\pmb{\mathscr{J}}$
are satisfied automatically if the evolution equations (\ref{eq:evolution1}) for $\mathscr{J}_{0}$ and the constraint equations (\ref{eq:constraints1}) are satisfied. This means the evolution equations (\ref{eq:evolution2}) for the vector component $\pmb{\mathscr{J}}$ are redundant.
As an example, we look at the first two equations of (\ref{eq:evolution2}).
It is easy to check that the first equation of (\ref{eq:evolution2}) holds following Eq. (\ref{eq:j0-j}).
Let us consider the second equation of (\ref{eq:evolution2}), in which we insert the second line of Eq. (\ref{eq:j0-j}) to obtain
\begin{equation}
G_0^{(0)}\pmb{\mathscr{J}}^{(0)} + \mathbf{G}^{(0)} \mathscr{J}_{0}^{(0)}
= \frac{1}{p_{0}} \mathbf{p} \times \left( \mathbf{G}^{(0)} \times \pmb{\mathscr{J}}^{(0)} \right) ,
\label{evo-derived}
\end{equation}
which is another evolution equation in the zeroth order in addition to the original one in (\ref{eq:evolution1}).
Using the first equation of (\ref{eq:j0-j}), the mass-shell condition in the zeroth order in (\ref{eq:constraints1}),
and the zeroth order evolution equation in (\ref{eq:evolution1}), one can verify that Eq. (\ref{evo-derived}) is automatically satisfied.
In the same way, one can verify that the last two equations in (\ref{eq:evolution2}) are also satisfied provided
Eqs. (\ref{eq:evolution1},\ref{eq:constraints1},\ref{eq:j0-j}) hold.

We have proved in the semiclassical expansion
that only $\mathscr{J}_0^{(i)}$ are independent and $\pmb{\mathscr{J}}$ can be expressed
by all $\mathscr{J}_0^{(i)}$ up to that order.
We can solve $\mathscr{J}_0^{(i)}$ order by order through the mass-shell conditions
which are obtained by substituting (\ref{eq:j0-j}) into (\ref{eq:constraints1}).
Then we can find the solutions for $\mathscr{J}_{0}^{(0,1,2)}$ as
\begin{eqnarray}
\mathscr{J}_{0}^{(0)} & = & p_{0}f^{(0)}\delta(p^{2}),\nonumber \\
\mathscr{J}_{0}^{(1)} & = & p_{0}f^{(1)}\delta(p^{2})+sQ(\mathbf{p}\cdot\mathbf{B})f^{(0)}\delta^{\prime}(p^{2}),\nonumber \\
\mathscr{J}_{0}^{(2)} & = & p_{0}f^{(2)}\delta(p^{2})+sQ(\mathbf{p}\cdot\mathbf{B})f^{(1)}\delta^{\prime}(p^{2})
+Q^2\frac{(\mathbf{p}\cdot\mathbf{B})^2}{2p_0}f^{(0)}\delta^{\prime\prime}(p^{2}) \nonumber\\
& &+\frac{1}{4p^2}\mathbf{p}\cdot \left\{ \mathbf{G}^{(0)}\times \left[\frac{1}{p_0}\mathbf{G}^{(0)}
\times \left(\mathbf{p}f^{(0)}\delta(p^{2}) \right) \right]\right\}
-\frac{{p_0}}{p^2} \Pi_{\mu}^{(2)}p^\mu f^{(0)}\delta(p^{2}) \nonumber\\
& &+\frac{1}{p^2} \mathbf{ p} \cdot \left( \mathbf{\Pi}^{(2)} p_0 - \Pi_{0}^{(2)} \mathbf{ p}  \right)f^{(0)}\delta(p^{2}) \,,
\label{eq:1st-order-sep-1}
\end{eqnarray}
where $f^{(0)}$, $f^{(1)}$ and $f^{(2)}$ are arbitrary scalar functions of $x$ and $p$
without singularity at $p^2=0$ and can be determined only by the original evolution equations in
(\ref{eq:evolution1}). Here we have used the derivative of a delta function
$\delta ^\prime (y)\equiv d\delta (y)/dy$. We can combine $\mathscr{J}_{0}^{(0)}$, $\mathscr{J}_{0}^{(1)}$
and the first three terms of $\mathscr{J}_{0}^{(2)}$ to obtain
\begin{equation}
\mathscr{J}_{0}\approx p_{0}f (x, p)\delta \left( p^{2}+\hbar sQ \frac{1}{p_{0}} \mathbf{p} \cdot \mathbf{B}  \right) ,
\end{equation}
where $f(x, p)\equiv f^{(0)}+\hbar f^{(1)}+\hbar^2  f^{(2)}$. We see that to the first order the energy poles
have been shifted to
\begin{equation}
E_p^{(\pm)}=\pm E_p (1 \mp \hbar sQ \mathbf{B}\cdot \boldsymbol{\Omega}_p),
\end{equation}
where $E_p=|\mathbf{p}|$ is the energy of the free fermion and
$\boldsymbol{\Omega}_{p}\equiv\mathbf{p}/(2|\mathbf{p}|^{3})$ is the Berry curvature in momentum space.
The energy correction can be regarded as the magnetic moment energy of
chiral fermions \cite{Son:2012zy,Manuel:2013zaa,Gao:2015zka}.
So to the first order the on-shell condition is modified by the magnetic moment energy,
but this does not work at the second order due to the last three terms of $\mathscr{J}_{0}^{(2)}$
which give the rest $O(\hbar ^2)$ contributions. The evolution equations (\ref{eq:evolution1})
can also be combined to give the CKE in four-momentum up to $O(\hbar ^2)$,
\begin{equation}
p_{\mu} G^{\mu}_{(0)}  \left[ f \delta( \tilde{p}^2 ) \right]
+ \frac{\hbar s}{2}\mathbf{G}^{(0)}\cdot\left\{\frac{1}{p_0}\mathbf{G}^{(0)}\times
\left[ \mathbf{p} f \delta ( \tilde{p}^2 )  \right]\right\}
+\hbar^2 C(f) =0 \,,
\label{eq:evolution1a}
\end{equation}
where the first two terms give the contribution up to the first order in $\hbar $ and
also part of the second order contribution, and $C(f)$ denotes the rest second order contribution given by
\begin{eqnarray}
C(f)&=&\frac{1}{4}\left(p^\mu G_{\mu}^{(0)}\frac{\mathbf{p}}{p^2}+\mathbf{G}^{(0)}\right)
\cdot \left[\frac{1}{p_0} \mathbf{G}^{(0)}\times \left(\frac{1}{p_0}\mathbf{G}^{(0)}
\times \mathbf{p}f \delta\left(p^{2}\right) \right)\right]\nonumber\\
& &+\left(p^\mu G_{\mu}^{(0)}\frac{\mathbf{p}}{p^2}+\mathbf{G}^{(0)}\right)
\cdot \left[\frac{1}{p_0}\left( \mathbf{\Pi}^{(2)} p_0 - \Pi_{0}^{(2)} \mathbf{ p}  \right)f \delta\left(p^{2}\right) \right]\nonumber\\
& & - p^\mu G_{\mu}^{(0)}
\left[\frac{1}{ p^2} \Pi_{\nu}^{(2)} p^\nu f \delta\left(p^{2}\right) \right]
+ G_{\mu}^{(2)} p^\mu f \delta\left(p^{2}\right)\,.
\label{2nd-cf}
\end{eqnarray}
Unlike the first order term in Eq. (\ref{eq:evolution1a}) which is proportional to the helicity $s$,
there is no explicit helicity dependence in the second order corrections besides $f_s$.
The second and third lines of Eq. (\ref{2nd-cf}) contribute only for varying or inhomogeneous fields
and vanish in constant background fields, i.e.
only the first line of Eq. (\ref{2nd-cf}) survives in constant fields as the second order contribution to the CKE.
Actually the first line of Eq. (\ref{2nd-cf}) is nonvanishing even without background fields
if there is a gradient of the distribution function.

\section{Chiral kinetic equation in three-momentum and chiral anomaly}
\label{cke-3d}

To obtain the CKE in three-momentum, we perform the integration of the CKE in four-momemtum in Eq. (\ref{eq:evolution1a})
over $p_{0}$ from $-\infty$ to $\infty$. The contribution from the $(-\infty,0)$ part of the integral gives the CKE
for antiparticles while the $(0,\infty)$ part gives that for particles.
We will present here the CKE in three-momentum  to $O(\hbar)$ for particles.
Up to the first order, the explicit form of Eq. (\ref{eq:evolution1a}) reads
\begin{eqnarray}
\label{J-eq1-3d-f-1}
& &\left[p_0\left(\partial_t  +Q\mathbf{ E }\cdot \pmb{ \nabla}_{p} \right)+\mathbf{ p }\cdot \left( \pmb{\nabla}_x  +Q \mathbf{ E} \partial_{p_0}
+Q\mathbf{ B }\times \pmb{ \nabla}_p\right) \right] \left[f \delta\left(p^2 + \frac{s\hbar Q \mathbf{ p}\cdot{\mathbf B}}{p_0 }\right) \right]\nonumber\\
& &+\frac{s\hbar}{2}\left\{ \left( \pmb{\nabla}_x  + Q \mathbf{ E} \partial_{p_0} +Q\mathbf{ B}\times \pmb{ \nabla}_p\right)\times
\left[\frac{1}{p_0}\left( \pmb{\nabla}_x+ Q\mathbf{ E}\partial_{p_0} +Q \mathbf{ B} \times{ \pmb{\nabla}}_p \right)\right] \right\} \nonumber\\
&& \cdot \left[{\mathbf p}f^{(0)}\delta\left(p^2+\frac{s\hbar Q\mathbf{p}\cdot \mathbf{ B}}{p_0}\right)\right]=0.
\end{eqnarray}
After the integration of the above equation over $p_0$ in the range $(0,\infty)$, we obtain
the CKE in three-momentum ($|\mathbf{p}| \neq 0$) to $O(\hbar)$ for particles (not antiparticles)
which has been previously derived
\cite{Son:2012zy,Manuel:2013zaa,Manuel:2014dza,Hidaka:2016yjf,Huang:2018wdl},
\begin{eqnarray}
\left(1+\hbar sQ\boldsymbol{\Omega}_{p}\cdot\mathbf{B}\right)\partial_{t}f(x,E_{p},\mathbf{p})\nonumber \\
+\left[\mathbf{v}+\hbar sQ(\mathbf{E}\times\boldsymbol{\Omega}_{p})+\hbar sQ\frac{1}{2|\mathbf{p}|^{2}}\mathbf{B}\right]\cdot\pmb{\nabla}_{x}f(x,E_{p},\mathbf{p})\nonumber \\
+\left[Q\tilde{\mathbf{E}}+Q\mathbf{v}\times\mathbf{B}+\hbar sQ^{2}(\mathbf{E}\cdot\mathbf{B})\boldsymbol{\Omega}_{p}\right]
\cdot\pmb{\nabla}_{p}f(x,E_{p},\mathbf{p}) & = & 0
\label{eq:cke-1}
\end{eqnarray}
where $\tilde{\mathbf{E}}\equiv\mathbf{E}-Q^{-1}\pmb{\nabla}_{x}E_{p}^{(+)}$ is the effective electric field,
$\mathbf{v}\equiv\pmb{\nabla}_{p}E_{p}^{(+)}$ is the effective velocity,
and $f(x,E_{p},\mathbf{p})$ is the distribution function on the mass-shell.
It should be noted that in the integration over $p_0$ from $0$ to $\infty$
some infrared singular terms emerge from the derivative in $p_0$ which cannot be dropped causally.
It turns out that there are two additional terms in the above CKE which are singular at $|\mathbf{p}| = 0$
but were previously neglected,
\begin{equation}
\hbar s \left(\mathbf{ E} \cdot \mathbf{B} \right)\left(\pmb{\nabla}_p\cdot \mathbf{ \Omega}_p \right) f(x,E_{p},\mathbf{p})
-\mathrm{lim}_{\Lambda\rightarrow 0} \frac{2\hbar s }{\Lambda} \left(\mathbf{ E}\cdot \mathbf{ p} \right)\left(\mathbf{ B} \cdot \mathbf{ p} \right)
\delta ^{\prime} (\Lambda^2- \mathbf{ p}^2 ) f (x,\Lambda ,\mathbf{p}) ,
\label{new-terms}
\end{equation}
where $\Lambda$ is an infrared cutoff introduced to regularize the integral over $p_0$ at $p_0=0$.
The last term comes from total derivatives in $p_0$ before integration over $p_0$.
We show the cutoff $\Lambda$ explicitly in order to emphasize its singularity nature.
Actually this term can be finally regularized into the term proportional
$\delta''(|\bf p|)$ and the dependence on $\Lambda$ will disappear eventually.
Note that these two terms in (\ref{new-terms}) only exist at $\mathbf{p}=\mathbf{0}$ or
in the deep infrared region of momentum. In this region, the kinetic description of chiral fermions
or the semiclassical expansion may not be valid.
In this case we need to consider all quantum contributions including the quantum mechanical description
of the particle motion \cite{Stephanov:2012ki}.

Although these new terms do not contribute to the chiral kinetic equation at finite momentum,
it is present in the anomalous conservation equation which is derived by integrating over the full momentum,
\begin{eqnarray}
\partial_{t}j_{0}+\pmb{\nabla}_x \cdot \mathbf{j} & = & -\frac{\hbar  s Q^2}{2}
 \int d^{3}\mathbf{p} \Big[ (\mathbf{E}\cdot\mathbf{B}) \boldsymbol{\Omega}_{p}\cdot \pmb{\nabla}_p f
+(\mathbf{E}\cdot\mathbf{B})\left(\pmb {\nabla}_p \cdot \boldsymbol{\Omega}_{p}\right) f \nonumber\\
&&- \mathrm{lim}_{\Lambda\rightarrow 0} \frac{ 2}{\Lambda} (\mathbf{E}\cdot\mathbf{p})(\mathbf{B}\cdot\mathbf{p})
\delta^{\prime}(\Lambda^2-\mathbf{ p}^2 ) f \Big] ,
\label{eq:anomalous-cons}
\end{eqnarray}
where the first term inside the square brackets in Eq. (\ref{eq:anomalous-cons})
is what is obtained before, the last two terms are related to those in Eq. (\ref{new-terms}).
The contributions to the chiral anomaly from the first and last term are identical at the limit $\Lambda \rightarrow 0$,
while the second term gives the same magnitude but the different sign.
Of course the last two terms cancel at the limit $\Lambda =0$, but we can see it differently:
the first two terms combine into a total divergence and vanish after the integration over $\mathbf{p}$,
while the last term contributes to the anomaly.
This implies that the chiral anomaly may arise from another source other than
the well-known Berry phase in three-momentum,
which seems to be consistent with the observation of Ref. \cite{Mueller:2017lzw}.

\section{Wigner functions in a general Lorentz frame}

In Sect. (\ref{semi-class}-\ref{cke-3d}), we work in a specific Lorentz frame.
It is easy and straightforward to rewrite all formula in a general Lorentz frame.
To this end, we need to introduce a time-like 4-vector $u^\mu$ with normalization $u^2=1$.
In general, $u^\mu$ can depend on space-time coordinates, but for simplicity we assume $u^\mu$ is a constant vector.
The purpose of this section is to show how to formulate our approach in a Lorentz covariant form,
we will only consider the contributions up to the first order of $\hbar$.
With the auxiliary vector $u^\mu$,  we can always decompose any vector $X^\mu$ into
the component parallel to $u^\mu$ and that perpendicular to $u^\mu$,
\begin{equation}
X^\mu=(X\cdot u) u^\mu + \bar X^\mu,
\label{decomp-u}
\end{equation}
with $\bar X \cdot u=0$. For $u^\mu = (1,0,0,0)$ we have the normal decomposition:
$X\cdot u=X_0$ and $\bar X^\mu=(0,\mathbf{X})$.
In such a decomposition, we can rewrite the Wigner equations (\ref{eq:constraint-1}-\ref{eq:constraint-2})
or (\ref{eq:evolution1}-\ref{eq:constraints2}) at the zeroth order as
\begin{eqnarray}
\label{J-eq-1-0}
u\cdot\nabla \left(u\cdot \mathscr{J}^{(0)} \right) +\bar \nabla\cdot  \bar{\mathscr{J}}^{(0)}&=& 0,\\
\label{J-c2-1b-0}
 \bar p_\mu \bar{\mathscr{J}}_\nu^{(0)}- \bar p_\nu \bar{\mathscr{J}}_\mu^{(0)}&=&0,\\
\label{J-c1-1-0}
(u\cdot p)(u\cdot \mathscr{J}^{(0)})+{\bar p}\cdot \bar{\mathscr{J}}^{(0)}  &=&0,\\
\label{J-c2-1a-0}
\bar p_\mu (u\cdot {\mathscr{J}}^{(0)}) - (u\cdot p) \bar{ \mathscr{J}}_\mu^{(0)} &=& 0 ,
\end{eqnarray}
and the equations of the first order as
\begin{eqnarray}
\label{J-eq-1-1}
u\cdot\nabla \left(u\cdot \mathscr{J}^{(1)} \right) +\bar \nabla\cdot  \bar{\mathscr{J}}^{(1)}&=& 0,\\
\label{J-c2-1b-1}
2s\left( \bar p_\mu \bar{\mathscr{J}}^{(1)}_\nu- \bar p_\nu \bar{\mathscr{J}}^{(1)}_\mu\right)&=&-
 \epsilon_{\mu\nu\rho\sigma}u^\rho  \left[ (u\cdot \nabla)\bar{ \mathscr{J}}^{(0)\sigma} - \bar\nabla^\sigma( u\cdot{\mathscr{J}}^{(0)})  \right] , \\
\label{J-c1-1-1}
(u\cdot p)(u\cdot \mathscr{J}^{(1)})+{\bar p}\cdot \bar{\mathscr{J}}^{(1)}  &=&0,\\
\label{J-c2-1a-1}
2s\left[ \bar p_\mu (u\cdot {\mathscr{J}^{(1)}})
- (u\cdot p) \bar{ \mathscr{J}}^{(1)}_{\mu} \right] 
&=& -\epsilon_{\mu\nu\rho\sigma} u ^\nu \bar \nabla^\rho\bar{ \mathscr{J}}^{(0)\sigma} ,
\end{eqnarray}
where $\nabla^\mu \equiv G_{(0)}^\mu = \partial^\mu_x - Q F^{\mu\nu}\partial_\nu^p$.
Just as we have shown in Sect. (\ref{semi-class}) and Sect. (\ref{2nd}),
Eq. (\ref{J-c2-1b-0}) and (\ref{J-c2-1b-1}) are redundant.
From Eq. (\ref{J-c2-1a-0}) and Eq. (\ref{J-c2-1a-1}), we obtain
\begin{eqnarray}
 \bar{ \mathscr{J}}_\mu^{(0)} &=& \bar p_\mu \frac{u\cdot {\mathscr{J}}^{(0)}}{ u\cdot p} ,\\
 \bar{ \mathscr{J}}_\mu^{(1)} &=& \bar p_\mu\frac{u\cdot {\mathscr{J}}^{(1)} }{ u\cdot p} 
-\frac{s}{2(u\cdot p)} \epsilon^{\mu\nu\rho\sigma} u_\nu \bar \nabla _{\sigma} \bar{ \mathscr{J}}^{(0)}_{\rho} 
\nonumber\\
&=& \bar p_\mu\frac{u\cdot {\mathscr{J}}^{(1)} }{ u\cdot p}
 -sQB_\mu\frac{1}{u\cdot p} \left( \frac{u\cdot {\mathscr{J}}^{(0)}}{ u\cdot p}\right)
 -\frac{s}{2(u\cdot p)}\epsilon^{\mu\nu\rho\sigma} u _\nu \bar p_\rho \bar \nabla_\sigma
\left( \frac{u\cdot {\mathscr{J}}^{(0)}}{ u\cdot p}\right) ,
\label{j0-j1-cov}
\end{eqnarray}
where $B^\mu =\frac 12 \epsilon ^{\mu\nu\alpha\beta} u_\nu F_{\alpha\beta}$ is the magnetic field four-vector.
We see from Eq. (\ref{j0-j1-cov}) that only the time-like component $u\cdot {\mathscr{J}}^{(0)}$
is independent and spatial components depend on it. 
Using the mass-shell conditions (\ref{J-c1-1-0}) and (\ref{J-c1-1-1})
we obtain the general solution,
\begin{eqnarray}
\frac{u\cdot {\mathscr{J}}^{(0)}}{ u\cdot p} &=& f^{(0)} \delta\left(p^2\right),\\
\frac{u\cdot {\mathscr{J}}^{(1)}}{ u\cdot p} &=&f^{(1)} \delta\left(p^2\right) - sQ\frac{B\cdot p }{u\cdot p}
f^{(0)}  \delta'\left(p^2\right) .
\end{eqnarray}
It follows that
\begin{eqnarray}
{\mathscr{J}}_\mu^{(0)} &=& p_\mu f^{(0)} \delta\left(p^2\right), \label{j0-mu} \\
{\mathscr{J}}_\mu^{(1)} &=& p_\mu f^{(1)} \delta\left(p^2\right)
- sQ p_\mu \frac{B\cdot p }{u\cdot p} f^{(0)}  \delta'\left(p^2\right)\nonumber\\
& & -sQ\frac{B_\mu}{u\cdot p}  f^{(0)} \delta\left(p^2\right)
-\frac{s}{2(u\cdot p)}\epsilon_{\mu\nu\rho\sigma} u ^\nu  p^\rho  \nabla^\sigma f^{(0)} \delta\left(p^2\right). \label{j1-mu}
\end{eqnarray}
Summing them up gives rise to the VWC up to the first order,
\begin{eqnarray}
{\mathscr{J}}_\mu &=& {\mathscr{J}}_\mu^{(0)} + \hbar {\mathscr{J}}_\mu^{(1)} \nonumber\\
&\approx&\delta \left(p^2-\hbar sQ \frac{B\cdot p}{u\cdot p}\right)
\left[ p_\mu f -\hbar sQ \frac{B_\mu}{u\cdot p}f
- \hbar s\frac{1}{2(u\cdot p)}\epsilon_{\mu\nu\rho\sigma} u^\nu  p^\rho  \nabla^\sigma f\right] ,
\end{eqnarray}
with $f\equiv f^{(0)}+\hbar f^{(1)}$. Then the covariant chiral kinetic equation
up to the first order is given by
\begin{eqnarray}
\nabla_\mu \left[\delta \left(p^2-\hbar sQ \frac{B\cdot p }{u\cdot p}\right)
\left( p^\mu -\hbar sQ \frac{B^\mu}{u\cdot p} - \hbar s
\frac{1}{2(u\cdot p)}\epsilon^{\mu\nu\rho\sigma} u _\nu  p_\rho  \nabla_\sigma \right) f \right] = 0.
\end{eqnarray}
Now all the expressions are written in a general Lorentz frame with the observer's velocity $u^\mu$.
Setting $u^\mu=(1,0,0,0)$ leads to the comoving frame chosen in previous sections.

We can also choose any other velocity $u^{\prime}_\mu$ to make the decomposition in (\ref{decomp-u}).
Then ${\mathscr{J}}_\mu^{(0)}$ and ${\mathscr{J}}_\mu^{(1)}$ in Eqs. (\ref{j0-mu},\ref{j1-mu})
can be expressed with $u^{\prime}_\mu$,
\begin{eqnarray}
{\mathscr{J}}^{\prime\mu}_{(0)}&=& p^\mu \frac{u'\cdot {\mathscr{J}}_{(0)}}{ u'\cdot p},\\
{\mathscr{J}}^{\prime\mu}_{(1)} &=& p^\mu \frac{u'\cdot {\mathscr{J}}_{(1)}}{ u'\cdot p}
 - \frac{s}{2u'\cdot p}\epsilon^{\mu\nu\rho\sigma} u' _\nu    \nabla_\sigma {\mathscr{J}}^{(0)}_\rho
\end{eqnarray}
which have been rewritten  in different but equivalent form from Eqs. (\ref{j0-mu},\ref{j1-mu}) for convenience in the following.
We can show that ${\mathscr{J}}_\mu^{(0)}$ and ${\mathscr{J}}_\mu^{(1)}$
are independent of the choice of $u^\mu$.
We can easily check ${\mathscr{J}}_{(0)}^{\prime\mu}={\mathscr{J}}_{(0)}^{\mu}$ as
\begin{eqnarray}
\delta{\mathscr{J}}^\mu_{(0)}
&=& {\mathscr{J}}^{\prime\mu}_{(0)}-{\mathscr{J}}^\mu_{(0)} \nonumber\\
&=& p^\mu  \frac{\left(u\cdot p\right)\left(u'\cdot {\mathscr{J}}_{(0)}\right)- \left(u'\cdot p\right)\left(u\cdot {\mathscr{J}}_{(0)}\right)}
{ \left(u'\cdot p\right)\left( u\cdot p\right) }\nonumber\\
&=& p^\mu  \frac{u^\rho u'^\sigma \left(p_\rho {\mathscr{J}}^{(0)}_\sigma-p_\sigma {\mathscr{J}}^{(0)}_\rho \right)}
{ \left(u'\cdot p\right)\left( u\cdot p\right) }\nonumber\\
&=&0,
\end{eqnarray}
where we have used ${\mathscr{J}}^\mu _{(0)}\propto p^\mu$ in the second to last equality. We can also verify ${\mathscr{J}}_{(1)}^{\prime\mu}={\mathscr{J}}_{(1)}^{\mu}$. To this end, we evaluate the difference 
\begin{eqnarray}
\delta{\mathscr{J}}^\mu_{(1)}&=& {\mathscr{J}}^{\prime\mu}_{(1)} - {\mathscr{J}}^\mu_{(1)},\nonumber\\
&=&  - \frac{ s p^\mu \epsilon^{\lambda\nu\rho\sigma} u_\lambda u'_\nu \nabla_\rho {\mathscr{J}}^{(0)}_\sigma }
{2 \left(u'\cdot p\right)\left( u\cdot p\right) }
-\frac{s p^\lambda \epsilon^{\mu\nu\rho\sigma}\left[  u_\lambda \left( u' _\nu    \nabla_\sigma {\mathscr{J}}^{(0)}_\rho \right)
-u'_\lambda\left( u _\nu    \nabla_\sigma {\mathscr{J}}^{(0)}_\rho \right)\right] }
{2\left(u'\cdot p\right)\left(u\cdot p\right)}
\label{verify-j1},
\end{eqnarray}
where we have used Eqs. (\ref{J-c2-1b-1},\ref{J-c2-1a-1}) in obtaining the first term.
We can rewrite the numerator of the second term by interchanging indices in the summation,
\begin{eqnarray}
&& p^\lambda \epsilon^{\mu\nu\rho\sigma}
\left[ u_\lambda \left( u' _\nu    \nabla_\sigma {\mathscr{J}}^{(0)}_\rho \right)
-u'_\lambda\left( u _\nu    \nabla_\sigma {\mathscr{J}}^{(0)}_\rho \right)\right] \nonumber\\
&=& \left( p^\lambda \epsilon^{\mu\nu\sigma\rho}-p^\nu \epsilon^{\mu\lambda\sigma\rho} \right)
u_\lambda u^{\prime}_\nu \nabla_\rho {\mathscr{J}}^{(0)}_\sigma \,. 
\end{eqnarray}
So Eq. (\ref{verify-j1}) can be simplified as
\begin{eqnarray}
\delta{\mathscr{J}}^\mu_{(1)} &=& -\frac{s}{2 \left(u'\cdot p\right)\left( u\cdot p\right)}
\left( p^\mu \epsilon^{\lambda\nu\rho\sigma}+
p^\lambda \epsilon^{\mu\nu\sigma\rho}-p^\nu \epsilon^{\mu\lambda\sigma\rho} \right)
u_\lambda u^{\prime}_\nu \nabla_\rho {\mathscr{J}}^{(0)}_\sigma
\nonumber\\
&=&- \frac{s}{2 \left(u'\cdot p\right)\left( u\cdot p\right)}
\left( p^\sigma \epsilon^{\rho\mu\lambda\nu}
+ p^\rho \epsilon^{\mu\lambda\nu\sigma}
\right) u_\lambda u^{\prime}_\nu \nabla_\rho {\mathscr{J}}^{(0)}_\sigma
\nonumber \\
&=&0,
\end{eqnarray}
where we have used Eqs. (\ref{J-eq-1-0}-\ref{J-c2-1a-0}) and the identity
\begin{equation}
p^\mu \epsilon^{\lambda\nu\sigma\rho}+
p^\lambda \epsilon^{\nu\sigma\rho\mu}+p^\nu \epsilon^{\sigma\rho\mu\lambda}
+ p^\sigma \epsilon^{\rho\mu\lambda\nu} + p^\rho \epsilon^{\mu\lambda\nu\sigma}=0 \,.
\end{equation}
Note that the first term of Eq. (\ref{verify-j1}) gives the non-trivial transformation of
the first order distribution function $\frac{u\cdot {\mathscr{J}}_{(1)}}{u\cdot p}$
from the change of the observer's velocity
\begin{equation}
\delta \left(\frac{u\cdot {\mathscr{J}}_{(1)} }{u\cdot p}\right)
=- \frac{ s \epsilon^{\lambda\nu\rho\sigma} u_\lambda u^{\prime}_\nu
\nabla_\rho {\mathscr{J}}^{(0)}_\sigma } {2 \left(u^\prime \cdot p\right)\left( u\cdot p\right) },
\end{equation}
which is related to the side-jump term first proposed in the study of 
Lorentz invariance in chiral kinetic theory \cite{Chen:2014cla} and later verified 
in the method of quantum field theory \cite{Hidaka:2016yjf}.  
We see that the distribution function in our approach can be unambiguously defined 
at the level of quantum field theory whose transformation in different frame 
emerges in a transparent way.

\section{Summary}

The quantum kinetics of chiral fermions is described by the vector component
of the covariant Wigner function with chirality (VWC).
We propose a semiclassical expansion of the VWC in background electromagnetic fields
in the Planck constant $\hbar$.
This expansion is very general and does not require quasi-equilibrium conditions as
in our previous works. We have shown to any order of $\hbar$ that only the time-components
(can be regarded as effective distribution functions) of the VWC are independent,
while the spatial components can be derived explicitly.
We have further demonstrated to any order of $\hbar$ that
a system of the quantum kinetic equations
for multiple-components of VWC can be reduced to one chiral kinetic equation (CKE)
for the single-component distribution function.
These are remarkable properties of quantum kinetics of chiral fermions and
will significantly simplify the description and
simulation of chiral effects in heavy ion collisions and chiral materials
such as Dirac or Weyl semimetals.
We have also derived the CKE in four-momentum up to the second order of $\hbar$.
We found additional terms in the CKE to $O(\hbar)$
which take effects in the infrared regime of momentum
and can contribute to the chiral anomaly  in the CKE other than the well-known
Berry phase term. We also show our method can be generalized to any Lorentz frame
with a reference four-velocity in a transparent way.
The side-jump effect naturally emerges from the change of the first order distribution function
when one chooses a different four-velocity.

\textit{Acknowledgments.} QW thanks D. Kharzeev, H.C. Ren and I. Shovkovy
for insightful discussions at the Huada-QCD school (2017)
in Wuhan when this project was initiated.
The authors thank J.W. Chen and C. Manuel for helpful discussions.
JHG is supported in part by the Major State Basic Research Development
Program (973 program) in China under Grant No. 2014CB845406, the National Natural
Science Foundation of China (NSFC) under Grant No. 11475104,
the Natural Science Foundation of Shandong Province under the Grant No. JQ201601 and
Qilu Youth Scholar Project Funding of Shandong University.
ZTL is supported in part by NSFC under Grant No. 11675092.
QW is supported in part by the 973 program under Grant No. 2015CB856902 and 2014CB845402
and by NSFC under Grant No. 11535012. XNW is supported by NSFC under Grant No. 11521064
and US DOE under Contract Nos. DE-AC02-05CH11231.

\appendix

\section{Proof of the statement in Section II}
\label{proof-app}


We will give the key steps to prove the following statement in Section II: for any
$n$, Eq. (7) is automatically satisfied once the
evolution equation (6) and the mass-shell conditions in Eq. (8) are
satisfied with Eq. (9). This statement can be put in another way: in the system
of equations (6-9) for any $n$, Eq. (7) is redundant.
We will call it the lemma for later reference. It is convenient to rewrite
$G^{\mu}$ and $\Pi^{\mu}$ as
\begin{eqnarray}
G^{\mu} & = & \sum_{k=0}^{\infty}\hbar^{k}G_{(k)}^{\mu},\hspace{1cm}\Pi^{\mu}=\sum_{k=0}^{\infty}\hbar^{k}G_{(k)}^{\mu},
\end{eqnarray}
where the zeroth order operators are given by $G_{(0)}^{\mu}=\partial_{x}^{\mu}-F^{\mu\nu}\partial_{\nu}^{p}$
and $\Pi_{(0)}^{\mu}=p^{\mu}$. For higher orders with $k\ge1$, these
operators have the following forms
\begin{eqnarray}
G_{(k)}^{\mu} & = & -C_{k}\Delta^{k}F^{\mu\nu}\partial_{\nu}^{p},\hspace{2cm}\Pi_{(k)}^{\mu}=kC_{k}\Delta^{k-1}F^{\mu\nu}\partial_{\nu}^{p},
\end{eqnarray}
with the coefficients $C_k$ defined by
\begin{equation}
C_{k}=\frac{\left[1+(-1)^{k}\right](-1)^{k/2}}{2^{k+1}(k+1)!}.\label{eq:ck}
\end{equation}
Note that for convenience we have absorbed the charge $Q$ into the
field strength $QF^{\mu\nu}\rightarrow F^{\mu\nu}$ so that it does not
appear in all formulas. The coefficients (\ref{eq:ck}) are equivalent
to those in Eq. (5) of the manuscript since all odd $k$ terms vanish.
We write these operators in three-dimension forms,
\begin{eqnarray}
\Pi_{0}^{(0)} & = & p_0,\nonumber \\
\mathbf{\Pi}^{(0)} & = & \mathbf{p},\nonumber \\
\Pi_{0}^{(k)} & = & -kC_{k}\Delta^{k-1}\mathbf{E}\cdot\pmb{\nabla}_{p},\nonumber \\
\mathbf{\Pi}^{(k)} & = & kC_{k}\Delta^{k-1}\left(\mathbf{E}\partial_{p_{0}}+\mathbf{B}\times\pmb{\nabla}_{p}\right),\nonumber \\
G_{0}^{(0)} & = & \partial_{t}+\mathbf{E}\cdot\pmb{\nabla}_{p},\nonumber \\
\mathbf{G}^{(0)} & = & \pmb{\nabla}_{x}+\mathbf{E}\partial_{p_{0}}+\mathbf{B}\times\pmb{\nabla}_{p},\nonumber \\
G_{0}^{(k)} & = & C_{k}\Delta^{k}\mathbf{E}\cdot\pmb{\nabla}_{p},\nonumber \\
\mathbf{G}^{(k)} & = & C_{k}\Delta^{k}\left(\mathbf{E}\partial_{p_{0}}+\mathbf{B}\times\pmb{\nabla}_{p}\right).
\label{op-explicit}
\end{eqnarray}
For convenience we can rewrite Eqs. (6,7,8,9) as
\begin{eqnarray}
\sum_{k=0}^{n}\left[G_{0}^{(k)}\mathscr{J}_{0}^{(n-k)}+{\mathbf{G}}^{(k)}\cdot{\pmb{\mathscr{J}}}^{(n-k)}\right] & = & 0,
\label{J0-eq-3d-n}\\
\sum_{k=0}^{n}\left[G_{0}^{(k)}\pmb{\mathscr{J}}^{(n-k)}+{\mathbf{G}}^{(k)}{\mathscr{J}}_{0}^{(n-k)}\right] & = & 2s\sum_{k=0}^{n+1}{\mathbf{\Pi}}^{(k)}\times\pmb{\mathscr{J}}^{(n-k+1)}.
\label{Ji-eq-3d-n}\\
\sum_{k=0}^{n}\left[\Pi_{0}^{(k)}\mathscr{J}_{0}^{(n-k)}-{\mathbf{\Pi}}^{(k)}\cdot{\pmb{\mathscr{J}}}^{(n-k)}\right] & = & 0,
\label{J0-cn1-3d-n}\\
\sum_{k=0}^{n}{\mathbf{G}}^{(k)}\times\pmb{\mathscr{J}}^{(n-k)} & = & -2s\sum_{k=0}^{n+1}\left[{\mathbf{\Pi}}^{(k)}{\mathscr{J}}_{0}^{(n-k+1)}-\Pi_{0}^{(k)}\pmb{\mathscr{J}}^{(n-k+1)}\right],
\label{Ji-cn2-3d-n}
\end{eqnarray}
We call these equations the $n$-th order equations or of order $n$
although ${\mathscr{J}}_{0}^{(n-k+1)}$, $\pmb{\mathscr{J}}^{(n-k+1)}$,
$\Pi_{0}^{(n+1)}$ and ${\mathbf{\Pi}}^{(n+1)}$ appears in the right-hand
sides of Eq. (\ref{Ji-eq-3d-n}) and (\ref{Ji-cn2-3d-n}). According
to this definition, the equations of order $n-1$ can be obtained
by the replacement $n\rightarrow n-1$ in Eqs. (\ref{J0-eq-3d-n}-\ref{Ji-cn2-3d-n}).
From Eq.(\ref{Ji-cn2-3d-n}), we obtain
\begin{eqnarray}
\pmb{\mathscr{J}}^{(n+1)} & = & \frac{1}{p_0} \left(\frac{s}{2}\sum_{k=0}^{n}{\mathbf{G}}^{(k)}\times\pmb{\mathscr{J}}^{(n-k)}\right.\nonumber \\
 &  & \left.+\sum_{k=0}^{n+1}{\mathbf{\Pi}}^{(k)}{\mathscr{J}}_{0}^{(n-k+1)}-\sum_{k=1}^{n+1}\Pi_{0}^{(k)}\pmb{\mathscr{J}}^{(n-k+1)}\right).
 \label{Ji-cn2-3d-n-1}
\end{eqnarray}
Note that this equation is another form of Eq. (\ref{Ji-cn2-3d-n})
so it is of order $n$.

In the following, we will prove the lemma by the mathematical induction
method. It is straightforward to verify that the lemma holds for the
order $n=0,1,2$. Hence we assume Eq. (\ref{Ji-eq-3d-n}) holds up
to the order $n-1$ with $n\geq3$ and that Eqs. (\ref{J0-eq-3d-n},\ref{J0-cn1-3d-n},\ref{Ji-cn2-3d-n-1})
are satisfied up to the order $n$, we need to prove that Eq. (\ref{Ji-eq-3d-n})
holds for the order $n$.

Substituting the expression (\ref{Ji-cn2-3d-n-1}) into the right-hand side (RHS) of  Eq.(\ref{Ji-eq-3d-n}) gives rise to
\begin{eqnarray}
\label{Ji-eq-3d-n-right}
\textrm{RHS}
& = &-\frac{2s}{p_0 } \sum_{k=1}^{n+1} {\mathbf \Pi}^{(0)}\times \left( \Pi_0^{(k)} \pmb{\mathscr{J}}^{(n-k+1)}\right)
+2s\sum_{k=1}^{n+1} {\mathbf \Pi}^{(k)}\times \pmb {\mathscr{J}}^{(n-k+1)}\nonumber\\
&=&-\frac{2s}{p_0}\sum_{k=0}^{n+1}\sum_{l=1}^{n-k+1}\Pi_0^{(l)} \mathbf \Pi^{(k)} \times \pmb{\mathscr{J}}^{(n-k-l+1)}
+\frac{1}{p_0 }\sum_{k=0}^{n} \sum_{l=0}^{n-k}\mathbf \Pi^{(k)}\times
\left( {\mathbf G}^{(l)}\times \pmb{\mathscr{J}}^{(n-k-l)}\right)\nonumber\\
& &+\frac{2s}{p_0 } \sum_{k=0}^{n+1} \sum_{l=0}^{n-k+1} \mathbf \Pi^{(k)}\times {\mathbf \Pi}^{(l)} {\mathscr{J}}_0^{(n-k-l+1)}
-\frac{2s}{p_0 }\sum_{k=1}^{n+1} \left[ {\mathbf \Pi}^{(0)},  \Pi_0^{(k)}\right] \times \pmb{\mathscr{J}}^{(n-k+1)}\nonumber\\
& &+2s\sum_{k=1}^{n+1}\left[ \mathbf \Pi^{(k)}, \frac{1}{p_0 }  \right] \times p_0
 \pmb{\mathscr{J}}^{(n-k+1)}
\end{eqnarray}
Using following relations
\begin{eqnarray}
&&\sum_{k=0}^{n+1} \sum_{l=0}^{n-k+1} \mathbf\Pi^{(k)}\times {\mathbf \Pi}^{(l)} {\mathscr{J}}_0^{(n-k-l+1)}
=- \sum_{k=1}^{n+1} (k+1) k C_k \Delta^{k-1}\mathbf B {\mathscr{J}}_0^{(n-k+1)} , \\
&&\sum_{k=1}^{n+1} \left[ {\mathbf \Pi}^{(0)},  \Pi_0^{(k)}\right] \times \pmb{\mathscr{J}}^{(n-k+1)}
=\sum_{k=1}^{n+1} k C_k\Delta^{k-1}\mathbf E \times \pmb{\mathscr{J}}^{(n-k+1)}  \nonumber\\
&&-\sum_{k=1}^{n+1} k (k-1) C_k\Delta^{k-2}\left[\pmb\nabla_x \left(\mathbf E\cdot\pmb \nabla_p\right) \right] \times \pmb{\mathscr{J}}^{(n-k+1)} , \\
&&\sum_{k=1}^{n+1}\left[ \mathbf \Pi^{(k)}, \frac{1}{p_0 } \right]p_0  \pmb{\mathscr{J}}^{(n-k+1)}
=-\frac{1}{p_0} \sum_{k=1}^{n+1} k C_k \Delta^{k-1} \mathbf E \times \pmb{\mathscr{J}}^{(n-k+1)} \nonumber \\
&&-\frac{1}{p_0} \sum_{k=1}^{n+1} k(k-1)C_k\Delta^{k-2}\left(\partial_t \mathbf E \partial_{p_0}+ \partial_t \mathbf B
\times \pmb \nabla_p\right)\times \pmb{\mathscr{J}}^{(n-k+1)} ,
\end{eqnarray}
and the Maxwell's equation $\partial _t \mathbf{B} + \pmb{\nabla } \times \mathbf{E} =0$, we obtain
\begin{eqnarray}
\textrm{RHS}&=&\frac{2s}{p_0}\sum_{k=0}^{n+1}\sum_{l=1}^{n-k+1}\Pi_0^{(l)} \mathbf \Pi^{(k)} \times \pmb{\mathscr{J}}^{(n-k-l+1)}
+\frac{1}{p_0 }\sum_{k=0}^{n} \sum_{l=0}^{n-k}\mathbf \Pi^{(k)}\times
\left( {\mathbf G}^{(l)}\times \pmb{\mathscr{J}}^{(n-k-l)}\right)\nonumber\\
& &-\frac{2s}{p_0}\sum_{k=2}^{n+1} (k+1) k C_k \Delta^{k-1}\left(\mathbf B {\mathscr{J}}_0^{(n-k+1)}+\mathbf E\times \pmb{\mathscr{J}}^{(n-k+1)}\right) .
\end{eqnarray}
We further use the identity $C_{k}=-\frac{1}{4 k (k+1)}C_{k-2}$ and finally rewrite the RHS as
\begin{eqnarray}
\textrm{RHS}&=&-\frac{2s}{p_0}\sum_{k=0}^{n+1}\sum_{l=1}^{n-k+1}\Pi_0^{(l)} \mathbf \Pi^{(k)} \times \pmb{\mathscr{J}}^{(n-k-l+1)}
+\frac{1}{p_0}\sum_{k=0}^{n} \sum_{l=0}^{n-k}\mathbf \Pi^{(k)}\times
\left( {\mathbf G}^{(l)}\times \pmb{\mathscr{J}}^{(n-k-l)}\right)\nonumber\\
& &+\frac{s}{2p_0}\sum_{k=0}^{n-1}  C_k \Delta^{k+1}\left(\mathbf B {\mathscr{J}}_0^{(n-k-1)}+\mathbf E\times \pmb{\mathscr{J}}^{(n-k-1)}\right) .
\label{finalRHS}
\end{eqnarray}

We now work on the left-hand side (LHS) of Eq. (\ref{Ji-eq-3d-n}),
\begin{eqnarray}
\textrm{LHS}
&=&\frac{1}{p_0 }G_0^{(0)}\left[\sum_{k=0}^n {\mathbf \Pi}_{(k)} {\mathscr{J}}_0^{(n-k)}+\frac{s}{2} \sum_{k=0}^{n-1}  {\mathbf G}^{(k)}\times \pmb{\mathscr{J}}^{(n-1-k)}
-\sum_{k=1}^n \Pi_0^{(k)} \pmb{\mathscr{J}}^{(n-k)}\right]\nonumber\\
& &+ \sum_{k=1}^n G_0^{(k)}\pmb{\mathscr{J}}^{(n-k)} + \sum_{k=0}^n {\mathbf G}^{(k)} {\mathscr{J}}_0^{(n-k)}\nonumber\\
&=&\frac{1}{p_0 }\left(\sum_{k=0}^n {\mathbf \Pi}^{(k)}G_0^{(0)} {\mathscr{J}}_0^{(n-k)}
+\frac{s}{2} \sum_{k=0}^{n-1}  {\mathbf G}^{(k)}\times G_0^{(0)}\pmb{\mathscr{J}}^{(n-1-k)}
-\sum_{k=1}^n \Pi_0^{(k)} G_0^{(0)}\pmb{\mathscr{J}}^{(n-k)}\right)\nonumber\\
& &+\frac{1}{p_0 }\left(\sum_{k=0}^n \left[G_0^{(0)},{\mathbf \Pi}^{(k)} \right] {\mathscr{J}}_0^{(n-k)}
+\frac{s}{2} \sum_{k=0}^{n-1} \left[G_0^{(0)}, {\mathbf G}^{(k)}\right]\times \pmb{\mathscr{J}}^{(n-1-k)}
-\sum_{k=1}^n\left[G_0^{(0)}, \Pi_0^{(k)}\right] \pmb{\mathscr{J}}^{(n-k)}\right)\nonumber\\
& &+ \sum_{k=1}^n G_0^{(k)}\pmb{\mathscr{J}}^{(n-k)} + \sum_{k=0}^n {\pmb G}^{(k)} {\mathscr{J}}_0^{(n-k)} .
\label{Ji-eq-3d-n-left}
\end{eqnarray}
Using Eq. (\ref{J0-eq-3d-n}) for $k\le n$ and Eq. (\ref{Ji-eq-3d-n}) for $k\le n-1$ yields
\begin{eqnarray}
\text{LHS}&=&-\frac{2s}{p_0 }\sum_{k=1}^{n+1} \sum_{l=0}^{n-k+1} \Pi_0^{(k)}
{\mathbf \Pi}^{(l)}\times \pmb {\mathscr{J}}^{(n-k-l+1)}
+\frac{1}{p_0 }\sum_{k=0}^{n} \sum_{l=0}^{n-k}{\mathbf \Pi}^{(k)} \times
\left({\mathbf G}^{(l)}\times \pmb {\mathscr{J}}^{(n-k-l)}\right) \nonumber\\
& &+\frac{1}{p_0 }\sum_{k=0}^{n} \sum_{l=0}^{n-k}\left( {\mathbf \Pi}^{(l)}\cdot {\mathbf G}^{(k)}
-{\mathbf G}^{(k)} \cdot  {\mathbf \Pi}^{(l)} \right) \pmb {\mathscr{J}}^{(n-k-l)}
-\frac{1}{p_0 }\sum_{k=0}^n \sum_{l=0}^{n-k} \left[ {\mathbf \Pi}^{(k)}, \mathbf G_i^{(l)} \right]
\pmb {\mathscr{J}}_i^{(n-k-l)} \nonumber\\
& &+\frac{1}{p_0 }\sum_{k=0}^{n} \sum_{l=0}^{n-k}\left[{\mathbf G}^{(k)},{\mathbf \Pi}_i^{(l)} \right] \pmb {\mathscr{J}}_i^{(n-k-l)}
+\frac{1}{p_0 }\sum_{k=0}^{n} \sum_{l=0}^{n-k}\left[ {\Pi}^{(l)}_0, {\mathbf G}^{(k)}\right] {\mathscr{J}}^{(n-k-l)}_0\nonumber\\
& &-\frac{1}{p_0 }\sum_{k=0}^n \sum_{l=0}^{n-k}\left[ {\mathbf \Pi}^{(k)}, G_0^{(l)} \right] {\mathscr{J}}_0^{(n-k-l)}
-\frac{s}{2p_0 }\sum_{k=0}^{n-1} \sum_{l=0}^{n-1-k} \left[{\mathbf G}^{(k)}, G_0^{(l)}\right] \times \pmb{\mathscr{J}}^{(n-1-k-l)} \nonumber\\
& &+\frac{1}{p_0 }\sum_{k=0}^n \sum_{l=0}^{n-k}  \left[\Pi_0^{(k)}, G_0^{(l)}\right]\pmb{\mathscr{J}}^{(n-k-l)}
-\frac{s}{2p_0 }\sum_{k=0}^{n-1} \sum_{l=0}^{n-1-k} {\mathbf G}^{(k)}\times \pmb G^{(l)}{\mathscr{J}}_0^{(n-1-k-l)} .
\end{eqnarray}
We use following relations
\begin{eqnarray}
& &\sum_{k=0}^{n} \sum_{l=0}^{n-k}\left( {\mathbf \Pi}^{(l)}\cdot {\mathbf G}^{(k)}
-{\mathbf G}^{(k)} \cdot  {\mathbf \Pi}^{(l)} \right) \pmb {\mathscr{J}}^{(n-k-l)}=0, \\
& &\sum_{k=0}^n \sum_{l=0}^{n-k}\left( \left[ {\mathbf \Pi}^{(k)}, \mathbf G_i^{(l)} \right]-\left[{\mathbf G}^{(k)},{\mathbf \Pi}_i^{(l)} \right]\right) \pmb {\mathscr{J}}_i^{(n-k-l)}=0, \\
& &\sum_{k=0}^{n} \sum_{l=0}^{n-k}\left(\left[ {\Pi}^{(l)}_0, {\mathbf G}^{(k)}\right]
-\left[ {\mathbf \Pi}^{(k)}, G_0^{(l)} \right] \right) {\mathscr{J}}^{(n-k-l)}_0=0 , \\
&&\sum_{k=0}^{n-1} \sum_{l=0}^{n-1-k} \left[{\mathbf G}^{(k)}, G_0^{(l)}\right] \times \pmb{\mathscr{J}}^{(n-1-k-l)}
=-\sum_{k=0}^{n-1} C_k \Delta^{k+1} \mathbf E \times \pmb{\mathscr{J}}^{(n-1-k)} , \label{a21}\\
&&\sum_{k=0}^n \sum_{l=0}^{n-k}  \left[\Pi_0^{(k)}, G_0^{(l)}\right]\pmb{\mathscr{J}}^{(n-k-l)}=0 , \\
&&\sum_{k=0}^{n-1} \sum_{l=0}^{n-1-k} {\mathbf G}^{(k)}\times \mathbf G^{(l)}{\mathscr{J}}_0^{(n-1-k-l)}
=-\sum_{k=0}^{n-1} C_k \Delta^{k+1} \mathbf B {\mathscr{J}}_0^{(n-1-k)} ,
\label{a23}
\end{eqnarray}
where we have used the Maxwell's equation $\partial _t \mathbf{B} + \pmb{\nabla } \times \mathbf{E} =0$
to obtain Eqs. (\ref{a21},\ref{a23}). Finally we arrive at
\begin{eqnarray}
\textrm{LHS}
&=&-\frac{2s}{p_0 }\sum_{k=1}^{n+1} \sum_{l=0}^{n-k+1} \Pi_0^{(k)}
{\mathbf \Pi}^{(l)}\times \pmb {\mathscr{J}}^{(n-k-l+1)}
+\frac{1}{p_0 }\sum_{k=0}^{n} \sum_{l=0}^{n-k}{\mathbf \Pi}^{(k)} \times
\left({\mathbf G}^{(l)}\times \pmb {\mathscr{J}}^{(n-k-l)}\right) \nonumber\\
& &+\frac{s}{2p_0 }\sum_{k=0}^{n-1} C_k \Delta^{k+1}\left(\mathbf B  {\mathscr{J}}_0^{(n-1-k)}
+ \mathbf E \times \pmb{\mathscr{J}}^{(n-1-k)} \right) ,
\end{eqnarray}
which is exactly the same as the RHS in Eq. (\ref{finalRHS}).

\bibliographystyle{apsrev}
\addcontentsline{toc}{section}{\refname}\bibliography{ref}

\end{document}